\documentclass{article}
\usepackage[letterpaper,top=1in,bottom=1in,left=1in,right=1in,marginparwidth=1.75cm]{geometry}
\usepackage[backend=biber,style=nature]{biblatex}
\begin{document}

\begin{center}
    \Large
    \textbf{Environmental Considerations in the age of Space Exploration: the Conservation and Protection of Non-Earth Environments}
        
    \vspace{0.4cm}
    \normalsize    
    \vspace{0.4cm}
    \textbf{Monica R. Vidaurri{$^1$}, Alexander Q. Gilbert{$^2$}}\par
    
    {$^1$}\textit{Department of Earth and Planetary Sciences, Stanford University, Stanford, CA} \par
    {$^2$}\textit{Colorado School of Mines, Nuclear Innovation Alliance}
\end{center}

\begin{abstract}
This document is an abbreviated version of the law review, led by Alexander Q. Gilbert, entitled: ``Major Federal Actions Significantly Affecting the Quality of the Space Environment: Applying NEPA to Federal and Federally Authorized Outer Space Activities." Here, we discuss the future of the space environment, and how it is increasingly becoming a human environment with regard to continued robotic and human presence in orbit, planned and proposed robotic and human presence on bodies such as the Moon and Mars, planned space mining projects, the increase use of low-Earth orbit for communications satellites, and other human uses of space. As such, we must evaluate and protect these environments just as we do on Earth. In order to prioritize mitigating threat of contamination, avoiding conflict, and promoting sustainability in space, all to ensure that actors maintain equal and safe access to space, we propose applying the National Environmental Policy Act, or NEPA, to space missions. We put forward three examples of environmental best practices for those involved in space missions to consider: adopting precautionary and communicative structure to before, during, and after missions taking place off-world, environmental impact statements, and transparency in tools that may impact the environment (including radioisotope power sources, plans in case of vehicle loss or loss of trajectory, and others). For additional discussion related to potential space applications of NEPA, NEPA's statutory text, and NEPA's relation to space law and judicial precedent for space, we recommend reading the full law review: \url{https://environs.law.ucdavis.edu/volumes/44/2/Gilbert.pdf}.
\end{abstract}

\section{Introduction}\label{sec:intro}
Within the last decade, planetary science and planetary exploration has progressed tremendously, and what was previously thought of as science fiction is rapidly becoming our reality. These technological and scientific advances, now augmented by non-governmental actors of the commercial sector, are redefining how humans interact with space and how we perceive our rights to and obligations in space. Space-based conversations and actions now focus on long-term habitation on other worlds, satellite constellations, space resource mining, and more. Not only is the human interaction with the space environment changing but the space environment itself is aligning with the issues and working definitions of the human environment. Within the next decade, environmental issues such as pollution, sustainability, climate change, contamination, and cooperation with international actors will become increasingly prevalent in non-Earth environments. As such, emerging priorities must include: avoiding conflict by ensuring that we are cognizant of our robotic and human explorers and their potential impact to non-Earth ecosystems; minimizing threat and contamination to non-Earth ecosystems and protecting their environmental integrity; and prioritizing international collaboration, the peaceful uses of space, the notion that space is for all, and an ancestral global commons. \par

To accomplish this, the planetary science community should consider the following practices:
\begin{itemize}
    \item Effectively applying the National Environmental Policy Act, or NEPA, and its precautionary and communicative structure to planetary missions before, during, and after mission completion, even if NEPA and other environmental law requirements do not explicitly change to include the space environment
    \item Making environmental impact statements and environmental assessments, as mandated by NEPA where it applies, mandatory for all space faring missions (similar to what the Office of Planetary Protection, or OPP, accomplishes) and have robust environmental risk management and mitigation processes for off-world procedure, including in any Earth orbit
    \item Synthesizing all areas of mission planning where the issue of environmental impact may be applied. This includes but is not limited to: radioisotope power sources, OPP assessment and decontamination procedure, plans in case of vehicle loss or loss of control in orbit or on trajectory, and plans in case of crashing onto a planetary surface
\end{itemize} \par 

We wish to highlight the importance of individual and group-level effort in accomplishing these goals, as expanding environmental law into outer space is a massive undertaking that can be catalyzed by creating norms. Considering the space environment as an environment with potential to be permanently negatively affected by efforts that ignore environmental consciousness is not an attempt to introduce unnecessary mitigation to the mission design and implementation process. Rather, we believe it is necessary work to ensure that space remains a peaceful global commons. We also wish to state that the purpose of this paper is not to discourage space exploration as a whole. Considering human uses and impacts to the space environment before mission development, and implementing the subsequent environmental risk mitigation policies, opens the door to cooperative and sustainable practices that can help the space science and planetary science community achieve its goals while reducing risk of potential legal and environmental conflict with other space-faring entities. Including the private sector in these conversations and standards further universalizes a prioritization of mitigating risk while encouraging innovation simultaneously, and having all voices of space exploration present.

\section{Space as a human environment}\label{sec:env}
\par To begin, it is important to remember that humanity has utilized the space environment since its beginnings. Though the definition of where the Earth ends and space begins is under continued debate \autocite{davalos_international_2016}, celestial bodies and even extrasolar objects have been a source of navigation, scientific study, time telling and calendar creation, or cultural and religious purposes. Even today, space continues to play those roles, now with electronic navigation, weather tracking, science, and internet satellites as well as human orbital spaceflight (the International Space Station) and space tourism. Often, the focus on these technological uses of space ignore other uses: the cultural and religious use of the space environment. When human ashes were spread on the lunar surface \autocite{fletcher_burying_2005}, Indigenous communities who view the moon and celestial bodies as an integral part of their culture raised concerns and objections \autocite{volante_navajos_1998}. Even on Earth, there are sacred Indigenous lands that are currently being used for scientific purposes related to outer space, with examples including telescope sites at Kitt Peak and Mt. Graham in Arizona, and notably, the halted construction of the Thirty Meter Telescope at Mauna Kea, and which is a designated protected site and whose project managers even filed an environmental impact statement \autocite{university_of_hawaii_at_hilo_thirty_2010}, but are facing continued conflict due to the sacred site that this particular telescope is posed to be built. Thus, it is imperative that these voices, who have every right to access space and their uses of it, are not ignored. Any future (and existing) missions and projects, as it shapes space environments for human use, should not just include but center Indigenous voices to avoid such conflict. As is the case with these telescopes, even if there are acceptable norms such as paying rent to Indigenous communities, offering internships and positions within the project, there will still be conflict if the project does not include these voices from the beginning \autocite{kahanamoku_native_2020}. Though this paper will mainly discuss direct environmental impacts in space such as pollution and debris, we wish to stress, first and foremost, that this new human environment is an ancestral global commons. As such, we must ensure that our exploration is respectful and inclusive to all involved. Any norms we create in environmental policy, protection, and avoiding contamination can only be strengthened by also centering respect and inclusion in these norms for those that have been astronomers for centuries, and stewards of the land that make this exploration possible. This is because at the heart of ensuring environmental standards in space is cooperation, and by highlighting this, we ensure this environment does not become dominated by few. \par 

Next, we must also consider that the planetary bodies that are accessible – the Moon, Mars and all solar system planets, asteroids, the moons of other planets, and the orbits of these celestial bodies – are of finite space and resources. These celestial bodies possess environments that, despite centuries of theory, exploration, and modeling, still possess unknown characteristics and can potentially be greatly impacted by sustained operations. Without any prior consensus on protection of these environments, missions or activities by even a sole actor could contaminate these bodies to the point where they pose safety risks to other actors, or prevent others from exploring these bodies. Even non-permanent operations pose unknown risks when considering volatile release during descent and ascent, and how it disperses various surface regoliths \autocite{prem_evolution_2020}. As space actors plan permanent outposts on celestial bodies (such as the Artemis mission on the lunar surface), these environments, though considered alien today, will become environments utilized and impacted by humans, similar to our current impacts of the Earth environment. \par 

Even our Earth’s orbit is becoming increasingly cluttered with space debris – defunct human-made satellites, telescopes, and other space objects – which may prove dangerous to current orbital spacecraft that are still in commission \autocite{garcia_space_2015}. Many orbital missions throughout history have been launched with no plan for coming back down, thus creating the issue of space debris, or space junk, and adding to the issue of safety when launching new orbital projects and missions. In addition, the recent introduction of satellite constellations have proved detrimental to the cultural, scientific, artistic, and recreational observation of the night sky, which is a global commons for all humanity \autocite{venkatesan_impact_2020}. In particular, the Starlink constellation satellites have raised many legal questions about not just the space environment as a human environment, but have challenged existing federal rules and regulations regarding operations – and especially commercial operations – in space. More generally, proposed satellite megaconstellations present dangers to observational capabilities, as well as weather forecasting and storm tracking \autocite{hainaut_impact_2020,constancewalker_impact_2020,noauthor_future_2019}. However, the Federal Communications Commision (FCC), as well as many other regulatory and scientific agencies, do not consider such projects as part of environmental review.

\section{NEPA}\label{sec:NEPA}
\par At the heart of the Starlink controversy of the night sky, and the argument of space as a human environment, is the National Environmental Policy Act (NEPA). NEPA is a U.S. environmental law that requires federal agencies to assess and potentially mitigate the impacts to the human environment posed by their work, prior to carrying out the work \autocite{us_epa_what_2013}. NEPA requires review of “major federal actions which significantly affect the quality of the human environment” \autocite{hughes_national_1975}, which includes but is not limited to: permit applications, construction sites, reviewing existing procedures, container storage, waste management, wildlife, and pollution. Under NEPA, federal agencies also have their own individualized Categorical Exclusions (often referred to as CatEx), which are actions that are deemed to not have a significant effect on the environment, and therefore do not require environmental assessment. These exclusions often refer to actions such as trail maintenance at National Parks, transferring of personal property (such as work computers and cell phones), and routine facilities maintenance. Specifically to the Starlink example, the authorizing federal agency for Starlink – the FCC – has most of its day-to-day operations listed as a Categorical Exclusion. FCC operations that provoke NEPA review are communications tower construction and federally protected land, tower height, antenna or equipment volume, high intensity lighting, and radiofrequency exposure. In this case, the FCC approved the Starlink constellation, as they deemed the operation in Earth orbit did not explicitly fall under their NEPA-required assessments nor did it constitute the human environment, and therefore was a categorical exclusion. As a result, there was backlash from astronomers and the public alike, and SpaceX had to figure out how to mitigate the issues presented by Starlink while its satellites had already been established in orbit \autocite{foust_report_2020,constancewalker_impact_2020}. \par 

This is a lesson that can be learned and prevented by the planetary science community – both public and private – to consider such issues before they arise. \textbf{We encourage regular review of categorical exclusions and discourage the allowance of space activities as categorical exclusions, as we affirm that the space environment should receive the same protections as the human environment, and these protections should be reviewed and strengthened at regular intervals.} In the case of Starlink, things are even simpler: while the activities in question occurred in space, they clearly affected the human environment on Earth by impacting professional terrestrial astronomers and others who value the night sky. Consideration of space as a human environment opens the door to increased protections of said environment, and protecting rights to access this environment.

\section{Current applications of environmental consciousness in planetary science}\label{sec:apps}
\par Applying environmental review where agencies may not be mandated need not be an entirely new endeavor, especially for planetary scientists. There exist multiple environmentally-conscious impact assessments throughout the planetary science mission planning process that simply need review and synthesizing, and the current NEPA policies that exist within federal agencies may provide quick and uncomplicated adoption of language to include environmental impacts in space. \par 

To begin, NASA Procedural Requirement (NPR) 8553.1C lays out NASA’s Environmental Management System including examples of environmental aspect categories which requires NEPA-mandated review \autocite{nasa_npr_2020}. In addition, the NPR includes nonconformity and corrective action, emergency preparedness and response, internal auditing (by use of Environmental Functional Reviews), and compliance evaluation, all of which helps provide a well-rounded system of checks and balances to ensure compliance with NEPA standards at every aspect. Other federal agencies have similar policies with similar checks in place. Continuous review and updating of these policies and applying them to space activities  is necessary and not without an already existing policy foundation. \par 

Within this framework, planetary scientists can consolidate and include current and potentially NEPA-prompting activities, such as NASA’s design reviews and a majority of the functions of the Office of Planetary Protection (OPP). NASA’s design reviews involve a series of meetings and project assessments on the technical and programmatic accountability of the project, and must be completed in order to apply funding. In summary, the project is assessed by a number of experts in the discipline of the project, who then decide whether or not the project design is achievable within NASA standards. In some agencies, such as the FDA, design reviews often require an independent reviewer. The design review process involves multiple various in-depth reviews, including system requirements review, mission design, system design, production readiness, test readiness, and operational readiness reviews. In reviewing every aspect of the mission including its intended outcomes, it is not beyond the scope of these reviews to include potential environmental impacts; both pertaining to the Earth environment and non-Earth environments. This particular assessment is present in the work of the OPP. \par

NASA’s OPP is an advisory body which assists with planetary exploration missions, ensuring that these missions adhere to planetary protection standards. This has historically included mitigating contamination on other celestial bodies or on Earth in the case of sample return or crew return, and maximizing protection of potential extraterrestrial life, ensuring that any life forms that are found in a non-Earth environment were not brought there by human explorers \autocite{nasa_npr_2011,committee_on_the_review_of_planetary_protection_policy_development_processes_review_2018}. In order to properly advise a mission, the OPP uses COSPAR mission categories to determine which planetary protection measures are applicable to the mission destination. The categories range from I-V, with I being no planetary protection measures needed (bodies with no direct interest for the study of the origins of life), and V being the most strict measures (sample return missions) \autocite{kminek_cospars_2019}. These mission categories, and subsequently the planetary protection and decontamination procedures as advised by the OPP, are directly influenced by our knowledge of the environments of celestial bodies that are of interest to solar system exploration missions. Extending the use of this knowledge to assess the impact of a mission on a planetary body provides a more than adequate integration of environmental consciousness to planetary science missions.\par 

In addition to the processes described above, there exists another established environmental hazard framework, which includes interagency review, for missions that utilize Radioisotope Heater Units, Radioisotope Power Systems, or nuclear reactors. As part of NASA’s compliance with NEPA, NASA missions utilizing these power sources undergo review alongside experts from the Department of Energy, the Environmental Protection Agency, and any other relevant agencies \autocite{bennett_flight_1987}. Though not likely, the potential harm from these power systems on other worlds is not improbable, and in addition to other environmental concerns stated in this document, potential harm from power sources to non-Earth environments can and should be assessed across all sectors. Currently, such missions are only required to conduct NEPA analysis for the launch portion of the mission, potentially omitting the environmental impacts that could occur in space from a mishap. \par

Finally, and perhaps most importantly, the leadership of the planetary science community in operating space environmental assessments can normalize environmental assessment for commercial space applications. Proposed commercial space applications, particularly space mining, could have large and negative impacts on space environments and potentially even harm extraterrestrial life. In order to maximize the benefits from advancing technology while minimizing the damages that these actors cause to planetary bodies and thus the risks they pose to planetary science, a precedent of NEPA or NEPA-like reviews can implant planetary protection values throughout all human space activities. This certainly includes our discussion about early inclusion and centering of historically excluded voices in mission planning; creating respect and inclusion as a norm will only strengthen new environmental protection norms that, as is inherent when operating in space, are centered around cooperation. 

\section{Conclusion}\label{sec:conclusion}
\par In summary, we encourage the planetary science community to apply environmental assessments and impact statements to all space-faring missions, whether or not they physically land on another celestial body, and synthesize all components of a mission which can pose an environmental impact, be it on Earth or in space. Exciting new endeavors and ideas in space exploration, and participation from the commercial sector, further validate the space environment – one that is a global commons, and that humans have been utilizing for centuries – as an environment that should be approached by prioritizing conservation, environmental protection, and the right to access this environment. This includes assessing the potential impacts of planetary missions to the destination environment (including the impacts it may have to the Earth and its inhabitants), and centering the voices of Indigenous communities in the process. By having these conversations regarding environmental impacts and conservation efforts in off-world scenarios early into mission design and review, and by actively including all involved in the mission (even indirectly), these efforts may be normalized and expected, even if language in NEPA and other environmental law itself does not explicitly change to reflect this. In addition, early and inclusive practice of these efforts may prevent further controversies as technology continues to progress.

\printbibliography
\end{document}